\documentstyle[12pt]{article}
\let\section=\subsection     \let\subsection=\subsubsection                

\begin{document}
\input epsf
\begin{center}
   {\large \bf MULTIFRAGMENTATION OF NON-SPHERICAL NUCLEI : 
ANALYSIS OF CENTRAL Xe + Sn COLLISIONS AT 50 MeV$\cdot$A} \footnote{Talk given
at the XXVII International Workshop on Gross Properties of Nuclei and Nuclear
Excitation, Hirschegg (Austria), January 17 - 23, 1999}\\[2mm]
   A.~LE~F\`{E}VRE, M.~P{\L}OSZAJCZAK and V.D.~TONEEV \\[5mm]
   {\small \it Grand Acc\'{e}l\'{e}rateur National d'Ions Lourds (GANIL)\\ 
    CEA/DSM -- CNRS/IN2P3, BP 5027, F-14076 Caen Cedex 05, France \\[8mm] }
\end{center}

\begin{abstract}\noindent
   The influence of shape of expanding and rotating source on
various characteristics of the multifragmentation process is studied.
The analysis is based on the
extension of the statistical microcanonical multifragmentation
model. The comparison with the data is done for central $Xe+Sn$ collisions at
$50 A\cdot MeV$ as measured by INDRA Collaboration.
\end{abstract}

\vfill
\newpage

The multifragmentation process has been studied in a broad range
of bombarding energies and for various types of projectiles. 
The reaction mechanism is often
considered in terms of two-step scenario where the first, 
dynamical step results in the formation of thermalized source
which then, in the second step, decays statistically into light particles
and intermediate-mass fragments (IMF's). Assuming that 
the thermal equilibrium is attained, various statistical
multifragmentation models were employed for the second step
(see \cite{gross90,bond96} and references quoted therein). These models
were so successful  that deviations between
their predictions and the experimental data have been often taken as an
indication for dynamical effects in the multifragmentation. However, 
one should be aware of  several oversimplifying assumptions in the
statistical calculations, such as , {\it e.g.},
the spherical shape of the thermalized source.
Indeed, one expects that the spherical shape is
perturbed during the dynamical phase and
 the density evolution can give
rise to complicated source forms \cite{GIT76}. Even more important are the
angular momentum induced shape instabilities
\cite{swiatecki} which may cause large fluctuations 
of both the Coulomb barrier
and the surface energy even for moderately high angular
momenta ($L \sim 40 \hbar$). 
In this paper, the non-spherical fragmenting source is considered within the
statistical model \cite{my}  which is based on the
MMMC method of the Berlin group \cite{gross90}. 
The observables sensitive to the source shape are
discussed and preliminary comparison with the experimental data for
$E_{kin}(Z)$ is presented. 
\begin{figure}[tbh]
\begin{center}
\leavevmode
\epsfxsize=14.3cm
\epsfbox{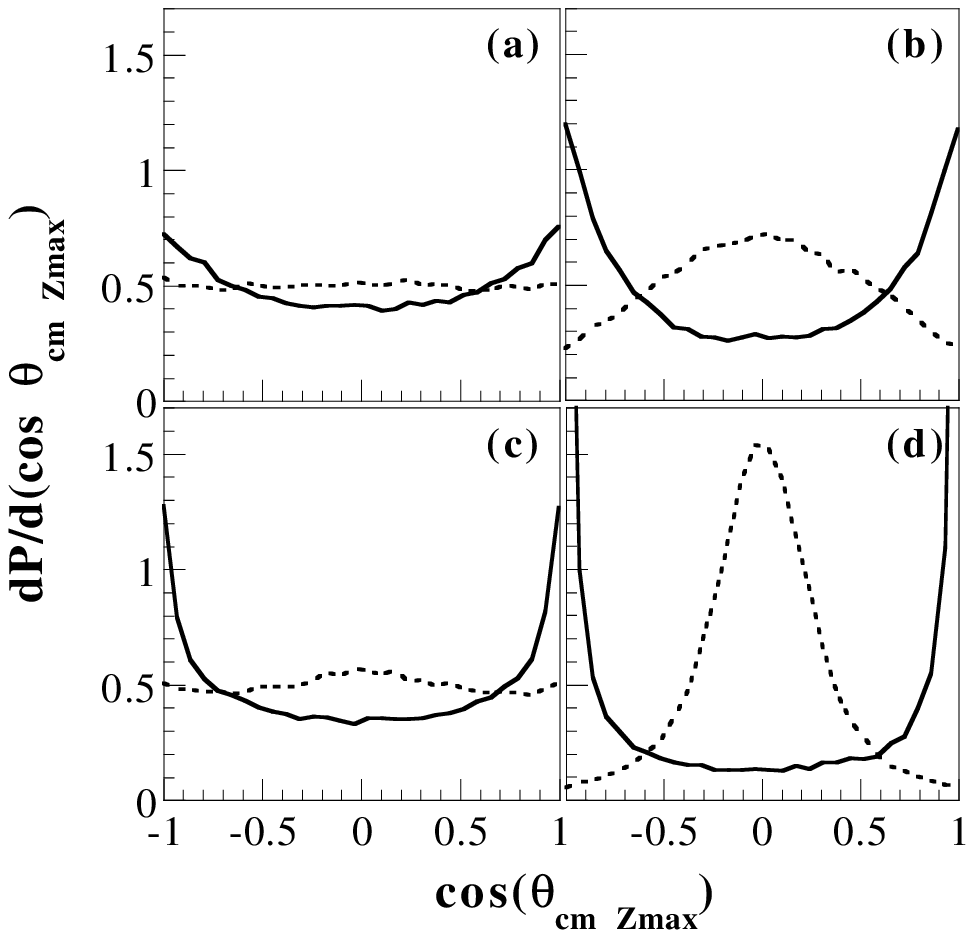}
\end{center}
{\begin{small} 
Fig.~1. Angular distribution of $Z_{max}$ in the c.m.s. Plots on 
the l.h.s. correspond to averaging over the whole available interval
$0\le\psi\le 2\pi$, whereas plots on the r.h.s. correspond to the 'frozen'
configuration  $ \psi = 0 $.                                 
{\bf (a),(b)}~: $L=40 \hbar$, $v_{\displaystyle b}=0$; \
{\bf (c),(d)}~: $L=40 \hbar$, $v_{\displaystyle b}=0.08 c$, $\alpha$=2 .  
\end{small}}
 \label{fig1}
\end{figure}

An explicit treatment of the fragment
positions in the occupied spatial volume allows for a
direct extension of the MMMC method 
to the case of non-spherical shapes \cite{my}. 
The source deformation in this case is
considered as an external constraint, similarly as the freeze-out volume. Below
we shall discuss axially symmetric ellipsoidal configurations
($R_x=R_y\neq R_z$) which are characterized by the ratio : 
${\cal R} = R_x/R_z$. The freeze-out volume of deformed system
is the same as that of an 'equivalent' spherical system with the radius
$R_{sys}=(R_xR_yR_z)^{1/3}$. Consequently, the statistical weights  
in the Metropolis scheme \cite{gross90} due to the 
accessible volume for fragments remain
the same.  The Coulomb energy in our model is  calculated exactly for   every
multifragment configuration of non-spherical nucleus. The general scheme to
account for the angular momentum in the calculation of statistical
weights is the same as discussed in Ref. \cite{BG95}. The source deformation
will change the moment of inertia of rotating system.  In calculating the 
statistical decay of 
fragmenting system, the angular velocity of the source is added to the 
thermal velocity of each fragment.
For each spatial configuration of fragments, part of the 
total energy goes into rotation and hence the temperature of the system will 
slightly fluctuate.  We take also into account fluctuations of the moment of 
inertia arising from fluctuations in the positions of fragments and 
light particles. 

In calculating all accessible states in the standard MMMC 
method \cite{gross90}, the source should be averaged with respect to
spatial orientations of its axes. Some of
these states will not be accessible 
if the angular momentum is conserved \cite{gross97,BG95} . We disentangle in
the MMMC code the beam direction (the $z$ -  axis)
and the rotation axis (the $x$ - axis, perpendicular to the reaction plane). 
The rotation energy is then :
${\bf L}^{2}/2J_x \equiv {L_x}^2/2J_x$, where
$J_x$ is the rigid-body moment of inertia
with respect to the $x$ axis.  Averaging
over polar angle $\theta$ is not consistent with the angular momentum
conservation. On the contrary, averaging over $2\pi$ in the angle $\phi$
corresponds to averaging over azimuthal angle of the reaction impact
parameter and should be included. Averaging over
rotation angle $\psi$ around ${\bf L}$ depends on the 
considered relationship between a rotation time :
$\tau_{rot}=J_x /L_x$, and a characteristic life-time of the source 
$\tau_{c}$. For high angular momenta when
$\tau_{rot}\ll\tau_{c} $, the full averaging in $0\le\psi\le 2\pi$ should be
performed.  In the opposite limit when $\tau_{rot}\gg\tau_{c} $, only states
with $\psi \simeq 0$ are accessible.  

In central HI collisions, part of the excitation energy can be stored in the
compression energy of pre-formed source which is transformed into the kinetic 
energy of fragments during the collective expansion . 
To get some insight into the influence  of collective flow on the
multifragmentation process,  we shall mimic this effect  by adding the
blast velocity $v_{\displaystyle b}$ to the thermal
velocity of each particle/fragment 
for any event simulated by the Metropolis method. The average collective energy
of expansion, similarly as the deformation energy,
is not included in the value of the total excitation energy. We
assume a simple scaling solution of non-relativistic hydrodynamics 
which yields the radial velocity profile \cite{BGZ78,CLZ94} :
\begin{equation}
\label{profile}
\label{bl} v_{\displaystyle b}(r) = v_0
\left({r}/{R_0} \right)^{\displaystyle \alpha} ~~~ \ ,
\end{equation}
where  $v_0$ and $R_0$ are the strength and scale parameters respectively, 
and the exponent $\alpha$ characterizes the power-low profile function. $R_0$
corresponds to the system size at the beginning of the scaling expansion
regime. Hence, $R_0<R_{sys}$ and we take $R_0=0.7R_{sys}$ in all studies.
Strictly speaking, the profile function (\ref{profile}) describes radial
expansion of a spherical source. For axially symmetric expansion, the velocity
profile may be more complicated. But even in this case, the scaling
solution (\ref{bl}) with  $0.5 \le \alpha \le 2$~
was successfully applied for describing the velocity profile
of the transverse expansion in high energy heavy-ion collisions  
\cite{CLZ94,PBM98,Heinz}. In the
multifragmentation case, we are dealing with
a non-spherical expansion of unstable matter and therefore
$\alpha$ may be treated as a free parameter.

Let us now consider the multifragmentation of $^{197}Au$ having the
angular momentum $L=40 \hbar$ and the thermal
excitation energy $6 \ A\cdot MeV$. 
These parameters characterize an
equilibrized source formed in central $Xe+Sn$ collisions at $50 \ A\cdot MeV$
studied by INDRA Collaboration \cite{bougault,INDRA}. 
All calculations are carried out at the
break-up density $\rho \approx \rho_0 /6$, what gives
$R_{sys}=12.8 \ fm$ for the effective radius of $^{197}Au$ source. 
In the following, we consider ellipsoidal forms with the ratio of axes 
${\cal R}=0.6$ (the prolate shape) and  ${\cal R}=1/0.6=1.667$ 
(the oblate shape).
We have found that none of the observables related to the fragment-size
distribution is sensitive to the source deformation at these
high excitation energies. On the contrary, 
the c.m.s. angular distribution of the largest fragment ($Z = Z_{max}$) 
is a highly sensitive observable (see Fig.1).  
In the absence of collective expansion,
the angular distribution of $Z_{max}$ 
is isotropic for oblate configurations and has small
forward - backward peaks for prolate configurations
if the $\psi$- averaging is performed in the whole available 
interval : $0\le\psi\le 2\pi$.
For the 'frozen' spatial configuration $( \psi = 0 )$, the shape effect is 
clearly seen: for the prolate form one finds strong forward - backward peaks,

\begin{figure}[tbh]
\begin{center}
\leavevmode
\epsfxsize=14.3cm
\epsfbox{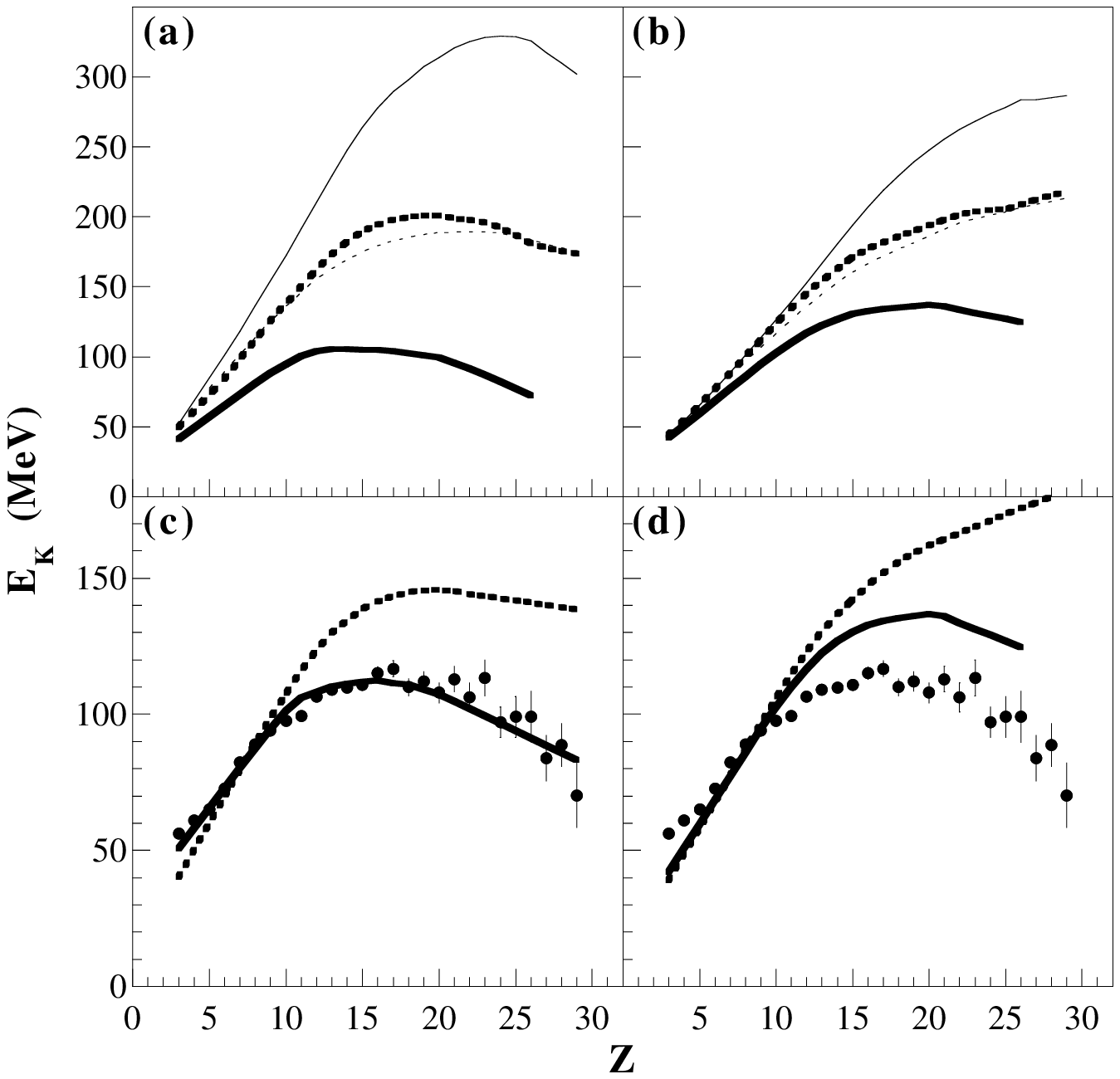}
\end{center}
{\begin{small} 
Fig.~2. Kinetic energy of IMF's is plotted as a function of $Z$. 
Thick lines correspond to events satisfying experimental
'centrality' condition : $\Theta_{flow} \geq \pi/3$ and 
accepted by the INDRA filter. Thin lines correspond to unfiltered events
and no condition on $\Theta_{flow}$. Prolate freezout configuration is
denoted with full lines and oblate configuration is denoted by the dashed
lines. The experimental data \cite{INDRA} are shown by dots. 
{\bf (a)} $L=40 \hbar$, $\alpha$=2, $v_{\displaystyle b}=0.08 c$. 
{\bf (b)} The same as in (a) but for $\alpha$=1.
{\bf (c)} Comparison between the experimental data and the calculation 
($\alpha$=2) for both prolate and oblate freeze-out configurations and the
blast velocities : $v_{\displaystyle b}=0.088 c$ (prolate shape), 
$v_{\displaystyle b}=0.063 c$ (oblate shape), chosen as to reproduce the 
experimental mean kinetic energy per nucleon of IMF's : $5.2 \pm 0.1 MeV/nucl$.
{\bf (d)} The same as in (c) but for $\alpha$=1. The blast velocities are :
$v_{\displaystyle b}=0.08 c$ and $v_{\displaystyle b}=0.07 c$ for prolate and
oblate freeze-out configuration respectively. \end{small} }
\label{fig2}
\end{figure}                
while for the oblate form the heaviest fragment is predominantly emitted  
in the sideward direction (${\theta}_{cm} = \pi /2$),
like in the hydrodynamic splashing scenario. 
The collective expansion ($\alpha = 2$) 
enhances the 'deformation effect'. One may notice a strong
increase of forward and backward peaks in the prolate case and the
appearance of a strong peak at  ${\theta}_{cm} = \pi /2$ in the oblate case.
Similar features can be seen also in the cumulative angular
distributions of all IMF's but the relative amplitude of the 'deformation 
effect' in that case is smaller.

Large sensitivity to the source shape can be seen in the analysis on an 
event-by-event basis using global variables, such as the
sphericity, coplanarity, aplanarity and the flow angle $\Theta_{flow}$
\cite{my}. In the latter case, the shape differences manifest
themselves even for $v_{\displaystyle b}=0$.
The whole effect is extremely sensitive to the collective expansion and 
is enhanced furthermore for the 'frozen' configuration ($\psi = 0$). It should
be mentioned that $\Theta_{flow}$-distribution  is the only observable
which explicitly depends on the angular momentum \cite{my}.

We have compared the model predictions with the experimental data
for central $Xe+Sn$ collisions at $50 \ A\cdot MeV$ 
\cite{bougault,INDRA,arnaud,salou}. 
Various statistical multifragmentation codes 
assuming the spherical source shape (the standard MMMC
\cite{gross90}, the SMM \cite{bond96}) 
have been tried earlier to explain these
data. All of them were successful in predicting fragment 
partitions \cite{bougault,arnaud,salou} which, as stated above, do not provide
a constraint on the source shape. On the other hand, these models failed to
reproduce the kinematical observables such as the IMF's angular 
distributions, the event shapes (sphericity, 
coplanarity, aplanarity), the IMF's 
average kinetic energies $E_{kin}(Z)$. In the latter case, 
the behaviour for heaviest fragments could not be reproduced even 
including radial expansion \cite{bougault}. 
To understand this issue, we have applied our model for different shapes, 
orientations, expansion profiles (following (1)), starting from the source of
$^{197}Au$ having $6 \ A\cdot MeV$ of 
thermal excitation energy and 
an angular momentum of $L=40 \hbar$. 
All calculated events have 
been filtered with the INDRA software replica, and then 
selected with the experimental centrality 
condition : complete events ({\it i.e.}, more than 80\% 
of the total charge and momentum is
detected) and $\Theta_{flow} \ge\ {\pi}/3$. We consider first 
the average kinetic energies of IMF's as a function of their charge. 
Results for different deformations ${\cal R}=0.6, 1.667$
and expansion parameterizations are compared in 
Fig. 2 with the experimental data \cite{bougault,INDRA,arnaud}. 
The calculations have been done for the 'frozen' configuration 
($\psi$=0). One can see that the data
{\it excludes} the radial velocity profile with $\alpha$=1 
and clearly {\it favours} the expanding prolate source with $\alpha$=2 and 
$\psi \sim$0. These source characteristics have been confirmed also by
the observables related to the event 
shape, like the  sphericity, coplanarity and aplanarity 
distributions, for which an excellent fit of the data 
has been obtained. Finally, we 
have examined the experimental IMF's and heaviest fragment 
angular distributions, as well as the $\Theta_{flow}$-distribution 
which again clearly discriminate between prolate, spherical 
and oblate source shapes.
On the basis of this whole set of observables, we believe to have the strong 
evidence in central $Xe+Sn$ collisions at $50 \ A\cdot MeV$ for the
formation of prolate, slowly rotating source with the orientation 
of the source aligned with the beam axis.

In conclusion, we have made an extension of the Berlin 
statistical multifragmentation code 
in order to include the effects of deformation, angular 
momentum and collective expansion. 
Due to the  change in the Coulomb energy for deformed freeze-out
configuration, the 'deformation effect' is clearly seen in the IMF's angular
distributions ($Z_{max}$- angular distribution) as well as
in the $\Theta_{flow}$ - distribution. 
A surprising interplay between effects of non-spherical 
freeze-out shapes and the memory effects of 
nonequilibrium phase of the reaction, such as the rotation and the collective
expansion of the source, has been found. 
The influence of shape on rotational properties of the 
system is not only reduced to the modification of the 
momenta of inertia. The limits on the averaging interval over
the angle $\psi$ about the rotation axis, which are defined by the 
time scales involved, affect strongly the angular
observables and are able  to enhance strongly the
'deformation effect'. These constraints may be important for certain
observables used in experimental procedures of selecting a
specific  class of events. Other striking finding is that the 
collective expansion allows to disclosure the source shape  in  the analysis 
 using global variables as well as
in the study of $Z$-dependence of $E_{kin}$. All these
experimental observables, including $E_{kin}(Z)$, are well reproduced assuming
strongly deformed (${\cal R}=0.6$), slowly rotating ($L=40 \hbar$) and
expanding source which has the radial velocity profile (1) with $\alpha \simeq
2$. Alongside with the observables discussed here, it would be interesting
to investigate the velocity and angular
correlations between fragments which are sensitive to the source shape at
the freeze-out. Such a work is now in progress \cite{future}. 

\vskip 1truecm
{\bf Acknowledgements} \\
We are grateful to D.H.E. Gross for his encouragement and interest 
in this project. We also thank G. Auger, A. Chbihi and 
J.-P. Wieleczko for their interest in this work.



\begin{thebibliography}{99}
\itemsep=0cm
\bibitem{gross90} D.H.E.~Gross, Rep.~Prog.~Phys. {\bf 53} (1990) 605.
\bibitem{bond96} J.P. Bondorf et al., Phys.~Rep. {\bf 257} (1996) 133.
\bibitem{GIT76} K.K.~Gudima et al., J.~Phys.~G: Nucl. and
Part.~Phys. {\bf 5} (1976) 229; W.~Bauer et al., Phys.~Rev.~Lett. {\bf
69} (1992) 1888; L.G.~Moretto et al.,   
Phys.~Rev.~Lett.  {\bf 69} (1992) 1884;
B.~Borderie et al., Phys. Lett. {\bf B 302} (1993) 15.
\bibitem{swiatecki} S.~Cohen and W.J.~Swiatecki, Annals~of~Physics (N.Y.) {\bf
82} (1974) 557; M.E.~Faber et al., Acta~Phys.~Pol. {\bf B15} (1984) 949.
\bibitem{my} A.~Le~F\`{e}vre, M.~Ploszajczak and V.D.~Toneev, nucl-th/9901099.
\bibitem{gross97} D.H.E.~Gross, Phys.~Rep. {\bf 279} (1997) 119.
\bibitem{BG95} A.S.~Botvina and D.H.E.~Gross, Nucl.~Phys.  {\bf A 592} (1995)
257.
\bibitem{BGZ78} J.P.~Bondorf, S.I.A.~Garpman and J.~Zim\'{a}nyi,
Nucl.~Phys. {\bf A 296} (1978) 320.
\bibitem{CLZ94} T.~Cs\"{o}rgo, B.~L\"orstad and J.~Zim\'{a}nyi, Phys.~Lett.
{\bf B 338} (1994) 134.
\bibitem{PBM98} A.~Polleri, J.P.~Bondorf and I.N.~Mishustin,  Phys.~Lett.
{\bf B 419} (1998) 19.
\bibitem{Heinz} K.S.~Lee and U.~Heinz,  Z. Phys. {\bf C43} (1989)
425;  Z. Phys. {\bf C48} (1990) 525.
\bibitem{bougault} R.~Bougault et al. (INDRA Collaboration), Proc. XXXV Int.
Winter Meeting on Nuclear Physics, ed. by I.~Iori , Bormio, (1997), p. 251.
\bibitem{INDRA} N.~Marie et al. (INDRA Collaboration),
Phys.~Lett. {\bf B 391} (1997) 15.
\bibitem{arnaud} A.~Le~F\`{e}vre, PhD-Thesis, GANIL T 97 03.
\bibitem{salou} S.~Salou, PhD-Thesis, GANIL T 97 06.
\bibitem{future} A.~Le~F\`{e}vre et al., in preparation.








\end{thebibliography}
\end{document}